# AC-Aware Steady-State Security Region Characterization via Exclusion Refinement

Zhirou Wang, *Student Member*, *IEEE*, Zhengshuo Li, *Senior Member*, *IEEE*, and Shuo Yang, *Student Member*, *IEEE*

***Abstract*—As active power dispatch becomes increasingly coupled with reactive power support and voltage security in modern power systems, AC security regions (AC-SRs) under exact steady-state AC power flow (AC-PF) constraints provide valuable insights into admissible operating boundaries and security margins for security assessment. However, the nonconvexity of exact AC-PF constraints makes it difficult to achieve both accuracy and computational efficiency in AC-SR characterization. To this end, this paper proposes an AC-aware iterative exclusion refinement (AC-ITER) approach that refines a given convex-relaxed AC-SR toward the exact AC-SR by progressively constructing and removing convex local exclusion subregions around numerically identified insecure points. Boundary-point searches and dual-based local exclusion cuts provide a tractable approximation of relaxation-induced insecure portions. A statistical certification method based on uniform sampling is developed to support high-confidence assessment of residual-region security and guide termination, while an AC-PF security similarity-based acceleration (AC-SSA) approach is introduced to further accelerate the sampling-based search for insecure operating points. Numerical studies on academic test systems and real-world systems demonstrate the characterization accuracy of AC-ITER and the computational efficiency gained through AC-SSA.**

## I. Introduction

### A. Background and Motivation

Security region (SR) can be typically interpreted as low-dimensional projections of high-dimensional security-constrained feasible sets onto selected operational variables [1], explicitly representing admissible operation boundaries together with their associated security margins. In bulk power systems, SRs associated with admissible power injection/exchange ranges have been widely studied for risk-aware operation and security assessment [2], [3]. To improve computational tractability, these SR studies commonly adopt DC power flow (DC-PF) models by neglecting reactive power and voltage constraints [2], [3], [4], under which SRs can be represented as convex polyhedra induced by linear constraint projections.

Driven by increasing renewable integration and expanding inter-regional power transfers, reactive power support and voltage security have become increasingly coupled with active power dispatch [5]. Recent investigations of the 2025 Iberian Peninsula blackout by the ENTSO-E Expert Panel [6] further highlighted the vulnerability of stressed systems to inadequate reactive support and voltage security violations associated with active power operating decisions. Under such conditions, SRs constructed under exact steady-state AC power flow (AC-PF) security constraints, referred to hereafter as AC security regions (AC-SRs), have become increasingly important for characterizing admissible active power operating ranges.

Unlike DC-based polyhedral SRs, AC-SRs induced by nonlinear AC-PF constraints may exhibit nonconvex and even disconnected geometries in low-dimensional projections [7], significantly complicating scalable SR characterization [8].

This work was supported in part by the National Natural Science Foundation of China (Project No. 52377107), in part by the Shandong Provincial Natural Science Foundation (Project ZR2026QB17), and in part by the Taishan Scholars Program. (*Corresponding author: Zhengshuo Li.*). The authors are with the School of Electrical Engineering, Shandong University, Jinan 250061, China.

### B. Literature Review

To explore the complex geometries of AC-SRs, several studies perform extensive feasible-point exploration under the exact AC-PF constraints. Common techniques include quotient gradient system (QGS)-based trajectory tracing [7], trajectory-unified (TJU) schemes [9], and Monte Carlo-style sampling that repeatedly solves AC optimal power flow (AC-OPF) problems to check the feasibility of candidate points [10]. Although these methods can provide valuable insights into the overall shape and connectivity of AC-SRs, they typically require a large number of trajectory integrations or repeated AC-OPF evaluations, becoming computationally intensive as system dimensionality and the number of sampled points increase.

Existing studies therefore mainly improve computational tractability from three perspectives:

**1) Geometry-oriented approximation**

Directly approximating AC-SRs through tractable geometric representations, including vertex and hyperplane search methods [11], [12], regular convex approximations such as hypercubes and ellipsoids [13], [14], data-driven boundary fitting techniques based on sampled feasible points [15], [16], and surrogate functions constructed to characterize the AC-SR boundaries [17]. These methods represent the AC-SR using simplified geometric boundaries or continuous surrogate functions but may fail to capture complex nonconvex geometries of the AC-SR, especially disconnected components and internal holes.

**2) Convexification of power flow equations**

Enhance SR tractability by replacing the nonconvex AC power flow equations with simplified or convexified formulations, including linearized DC power flow equations [2], [3] and convex relaxations based on SOCP/QCQP and semidefinite programming formulations [18], [19], [20], [21]. These methods characterize SR boundaries through duality-based cut generation and related decomposition approaches. With sufficiently refined cutting planes, the resulting convex polyhedra can numerically approximate the projected feasible set of the convexified power flow model. Consequently, relaxation-based formulations generally provide outer approximations of exact AC-SRs. Despite their computational tractability, the relaxation gaps and approximation errors introduced by convexified AC-PF models may become pronounced under stressed operating conditions with limited Volt/Var support, causing convex-relaxed SRs to include operating points that remain infeasible under the exact AC-PF security constraints [20], thereby leading to overestimated security margins.

**3) Post-relaxation refinement**

Refining SRs through post-construction correction offers another route to improved SR accuracy. Under DC-PF formulations with unit commitment, [4] applies Farkas-lemma-based exclusion cuts with logical constraints to iteratively identify and remove internal infeasible subregions caused by generator on/off decisions, yielding a tighter SR within the DC-PF model. Similarly, [18] employs heuristic elimination based on SOCP dual variables to prune inexact regions from a Distflow-based SR construction. These methods demonstrate the value of iterative correction within their respective relaxed SRs. However, the refinement in both cases is directed toward improving the fidelity of the DC-PF or Distflow SR to its own underlying model, not toward the exact AC-PF security constraints. Operating points that are infeasible under the exact AC-PF equations but admitted by these relaxed constructions therefore remain outside the scope of such corrections.

Across all three perspectives, accurate AC-SR characterization remains fundamentally constrained by the tradeoff between computational tractability and fidelity to the exact AC-PF model, leaving the elimination of operating points violating exact AC-PF security constraints from tractable SR representations as an open challenge.

### C. Contributions and Organization

To address the above challenges, this paper proposes an AC-aware iterative exclusion refinement (AC-ITER) approach for improving the fidelity of convex-relaxed AC-SRs with respect to the exact AC-SR. Finite-sample statistical guarantees and an approach to accelerate sampling-based AC-PF security assessment are further developed to support AC-ITER. The main contributions of this paper are summarized as follows:

A) An AC-ITER approach is proposed to refine convex-relaxed AC-SRs by progressively removing relaxation-induced insecure subregions. Since directly identifying such subregions requires solving a bilevel nonconvex program for which exact solutions are generally unattainable, AC-ITER introduces a statistical security criterion based on uniform sampling to guide the detection of insecure operating points and termination, and constructs convex local exclusion subregions around these points, thereby systematically improving the fidelity of convex-relaxed AC-SRs with respect to the exact AC-SR.

B) Finite-sample statistical guarantees are established for a fixed region using independently sampled points. When all sampled points are secure, the analysis determines how many samples are needed to meet the prescribed confidence level and security requirement. This sample requirement guides termination in AC-ITER, while previously verified secure samples are reused to reduce repeated assessments.

C) An AC-PF security similarity-based acceleration (AC-SSA) approach is developed to accelerate sampling-based AC-PF security assessment in AC-ITER. Building on the state-similarity-based approach in [22] and the parallel *hypothesis-then-verify* strategy in [2], AC-SSA assesses newly sampled points in AC-ITER by reusing information from nearby operating points already verified as secure, reducing the need to solve a nonlinear optimization problem for each sample. Combined with parallel processing, the proposed approach effectively improves the computational efficiency of AC-ITER.

The rest of this article is organized as follows: Section II formulates the definition of AC-SR and its convex relaxation. Section III presents the AC-ITER approach. Section IV establishes finite-sample statistical guarantees and the AC-SSA acceleration approach for AC-ITER. Case studies are carried out in Section V, followed by conclusions in Section VI.

## II. Problem Formulation of AC-SR

### A. Definition of AC-SR

Under the AC-PF formulation, steady-state power system security is characterized by bus power injections together with branch power flow constraints. Based on the definition of complex power as the product of bus voltage and the conjugate of injected current, the bus complex power injection vector $\tilde{\boldsymbol{f}}_{\text{inj}}$ together with the branch complex power flow vectors $\tilde{\boldsymbol{f}}_{\text{fr}}$ and $\tilde{\boldsymbol{f}}_{\text{to}}$ can be compactly expressed as:

$$\tilde{\boldsymbol{f}}_{\text{inj}}(\boldsymbol{\theta},\boldsymbol{U}) = \tilde{\boldsymbol{V}} \odot \text{conj}\left(\mathbb{Y}_{\text{bus}}\tilde{\boldsymbol{V}}\right) \in \mathbb{C}^{n\times 1} \tag{1}$$

$$\tilde{\boldsymbol{f}}_{\text{fr}}(\boldsymbol{\theta},\boldsymbol{U}) = \boldsymbol{C}_{\text{fr}}\tilde{\boldsymbol{V}} \odot \text{conj}\left(\mathbb{Y}_{\text{fr}}\tilde{\boldsymbol{V}}\right) \in \mathbb{C}^{l\times 1} \tag{2}$$

$$\tilde{\boldsymbol{f}}_{\text{to}}(\boldsymbol{\theta},\boldsymbol{U}) = \boldsymbol{C}_{\text{to}}\tilde{\boldsymbol{V}} \odot \text{conj}\left(\mathbb{Y}_{\text{to}}\tilde{\boldsymbol{V}}\right) \in \mathbb{C}^{l\times 1} \tag{3}$$

where $\odot$ denotes the Schur product, $n$ and $l$ correspond to the numbers of buses and branches, respectively. The complex voltage vector is given by $\tilde{\boldsymbol{V}} = \boldsymbol{U}e^{j\boldsymbol{\theta}}$, with $\boldsymbol{U}$ and $\boldsymbol{\theta}$ collecting the bus voltage magnitudes and phase angles, respectively. $\mathbb{Y}_{\text{bus}} \in \mathbb{C}^{n\times n}$ is the system bus admittance matrix, whereas $\mathbb{Y}_{\text{fr}}, \mathbb{Y}_{\text{to}} \in \mathbb{C}^{l\times n}$ characterize the branch admittances associated with the from- and to-ends of transmission lines. Moreover, $\boldsymbol{C}_{\text{fr}}, \boldsymbol{C}_{\text{to}} \in \{0,1\}^{l\times n}$ are the corresponding branch-bus incidence matrices for the from- and to-side connections.

In this paper, the net complex power injection $\tilde{\boldsymbol{S}}_{\text{inj}}$ at each bus is determined by the combined contributions of generators and loads, which can be compactly represented as:

$$\tilde{\boldsymbol{S}}_{\text{inj}} = \boldsymbol{C}_{\text{G}}\tilde{\boldsymbol{S}}_{\text{G}} + \tilde{\boldsymbol{S}}_{\text{L}} \in \mathbb{C}^{n\times 1}, \tag{4}$$

where the term $\tilde{\boldsymbol{S}}_{\text{L}} = \boldsymbol{P}_{\text{L}} + j\boldsymbol{Q}_{\text{L}}$ and $\tilde{\boldsymbol{S}}_{\text{G}} = \boldsymbol{P}_{\text{G}} + j\boldsymbol{Q}_{\text{G}}$ represent the complex power of the load, generator respectively, and $\boldsymbol{C}_{n\times g}$ is the generator-to-bus incidence matrix.

The feasible set $\mathcal{F}^{\text{AC}}_{\boldsymbol{X},\boldsymbol{Y}}$ of the security-constrained AC-PF model, characterized by the nonlinear power-balance equations together with the associated operational security limits, can be compactly written as:

$$\mathcal{F}^{\text{AC}}_{\boldsymbol{X},\boldsymbol{Y}} \triangleq \left\{ \begin{matrix} (\boldsymbol{X},\boldsymbol{Y}) \\ \mathcal{X}\times\mathcal{Y} \end{matrix} \middle| \begin{matrix} \boldsymbol{G}_{\text{inj}}(\boldsymbol{X},\boldsymbol{Y}) = \boldsymbol{0}, \\ \boldsymbol{H}_{\text{cap}}(\boldsymbol{X},\boldsymbol{Y}) \le \boldsymbol{0} \end{matrix} \right\} \tag{5}$$

$$\boldsymbol{G}_{\text{inj}}(\boldsymbol{X},\boldsymbol{Y}) = \begin{bmatrix} \text{Re}\left\{\tilde{\boldsymbol{f}}_{\text{inj}}(\boldsymbol{\theta},\boldsymbol{U}) - \tilde{\boldsymbol{S}}_{\text{inj}}\right\} \\ \text{Im}\left\{\tilde{\boldsymbol{f}}_{\text{inj}}(\boldsymbol{\theta},\boldsymbol{U}) - \tilde{\boldsymbol{S}}_{\text{inj}}\right\} \end{bmatrix} \in \mathbb{R}^{2n\times 1} \tag{6}$$

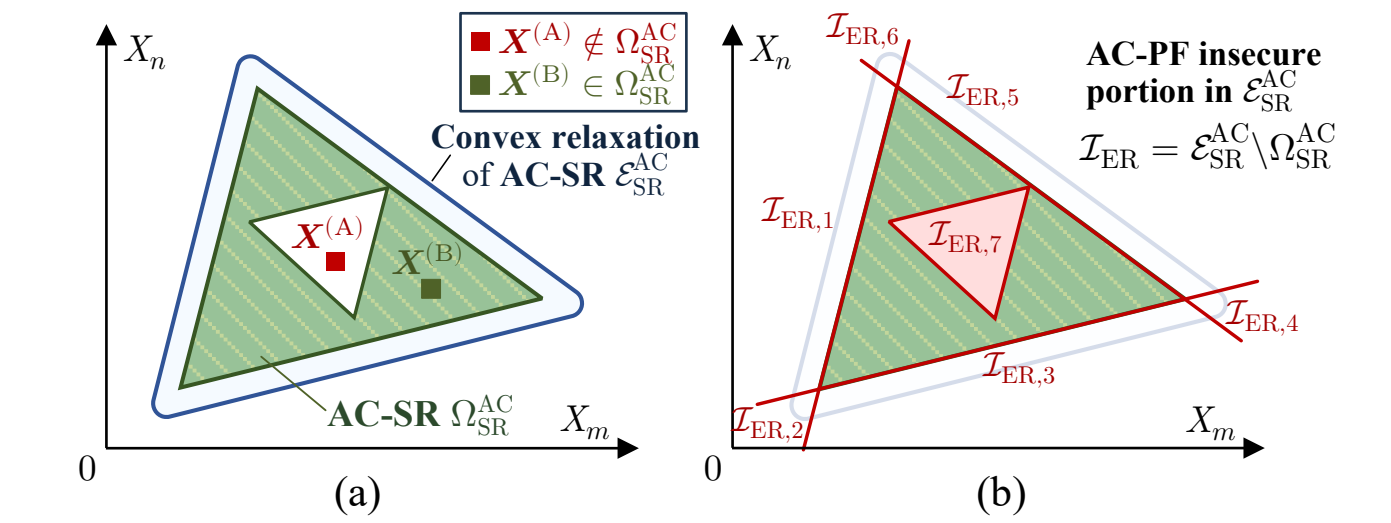

Fig. 1. (a) The convex relaxation $\mathcal{E}_{\mathrm{SR}}^{\mathrm{AC}}$ forms an outer relaxation of the exact AC-SR $\Omega_{\mathrm{SR}}^{\mathrm{AC}}$, admitting AC-PF insecure operating points (e.g. $\boldsymbol{X}^{(\mathrm{A})} \notin \Omega_{\mathrm{SR}}^{\mathrm{AC}}$) that overestimate the security margin; (b) The region to be excluded $\mathcal{I}_{\mathrm{ER}} = \mathcal{E}_{\mathrm{SR}}^{\mathrm{AC}} \backslash \Omega_{\mathrm{SR}}^{\mathrm{AC}}$, which AC-ITER covers by a union of convex polytopes $\{\mathcal{I}_{\mathrm{ER},k}\}_{k=1}^{7}$ and removes to refine the relaxation toward the exact AC-SR.

$$\boldsymbol{H}_{\mathrm{cap}}(\boldsymbol{X},\boldsymbol{Y}) = \begin{bmatrix} \left|\tilde{\boldsymbol{f}}_{\mathrm{fr}}(\boldsymbol{\theta},\boldsymbol{U})\right| - \overline{\boldsymbol{S}_{\mathrm{line}}} \\ \left|\tilde{\boldsymbol{f}}_{\mathrm{to}}(\boldsymbol{\theta},\boldsymbol{U})\right| - \overline{\boldsymbol{S}_{\mathrm{line}}} \\ \boldsymbol{AX}+\boldsymbol{BY}-\boldsymbol{D} \end{bmatrix} \in \mathbb{R}^{(4n+2l+4g)\times 1} \tag{7}$$

$$(\boldsymbol{X},\boldsymbol{Y}) = [\boldsymbol{\theta},\boldsymbol{U},\boldsymbol{P}_{\mathrm{G}},\boldsymbol{Q}_{\mathrm{G}}]^{\mathrm{T}} \in \mathbb{R}^{(2n+2g)\times 1} \tag{8}$$

where $\boldsymbol{X}$ collects the observed variables that define the SR, $\boldsymbol{Y}$ contains the remaining continuous operational variables, such as voltage phase angles $\boldsymbol{\theta}$ and voltage magnitudes $\boldsymbol{U}$. The sets $\mathcal{X}$, $\mathcal{Y}$ constrain the ranges of $\boldsymbol{X}, \boldsymbol{Y}$ respectively. The equality constraints $\boldsymbol{G}_{\mathrm{inj}}(\cdot,\cdot) = \boldsymbol{0}$, defined in (6), enforce nodal active and reactive power balance equations, while the inequality constraints $\boldsymbol{H}_{\mathrm{cap}}(\cdot,\cdot) \le \boldsymbol{0}$, defined in (7), represent transmission and operational security limits. The affine term $\boldsymbol{AX}+\boldsymbol{BY}-\boldsymbol{D}$ collects the linear operational constraints on $\boldsymbol{X}$ and $\boldsymbol{Y}$, including upper and lower bounds on decision variables[1].

**Definition 1** (*AC-PF Security of an Operating Point*): An observed operating point $\boldsymbol{X} \in \mathcal{X}$ is said to be **AC-PF secure** if there exists at least one complementary state vector $\boldsymbol{Y} \in \mathcal{Y}$ such that $(\boldsymbol{X},\boldsymbol{Y}) \in \mathcal{F}_{\boldsymbol{X},\boldsymbol{Y}}^{\mathrm{AC}}$. i.e., the full system state $(\boldsymbol{X},\boldsymbol{Y})$ simultaneously satisfies all AC power balance equations (6) and security inequality constraints (7). Otherwise, the operating point is referred to as **AC-PF insecure**.

**Definition 2** (*AC Security Region*): The *AC Security Region* (AC-SR) is defined on the observed variables $\boldsymbol{X}$ as the projection of the $\mathcal{F}_{\boldsymbol{X},\boldsymbol{Y}}^{\mathrm{AC}}$. i.e. the set of all $\boldsymbol{X}$ for which there exists $\boldsymbol{Y} \in \mathcal{Y}$ satisfying the security-constrained AC-PF model:

$$\Omega_{\mathrm{SR}}^{\mathrm{AC}} \triangleq \left\{\boldsymbol{X} \middle| \exists \boldsymbol{Y} \in \mathcal{Y}, (\boldsymbol{X},\boldsymbol{Y}) \in \mathcal{F}_{\boldsymbol{X},\boldsymbol{Y}}^{\mathrm{AC}}\right\} \tag{9}$$

i.e., the set of operating conditions on $\boldsymbol{X}$ for which at least one AC-PF solution satisfying all security constraints exists. Accordingly, the definition of AC-SR $\Omega_{\mathrm{SR}}^{\mathrm{AC}}$ is precisely the set of all AC-PF secure operating points under the security-constrained model. Owing to the nonlinear AC-PF equations and operational inequality constraints embedded in $\mathcal{F}_{\boldsymbol{X},\boldsymbol{Y}}^{\mathrm{AC}}$, the resulting AC-SR is generally non-convex and may even become disconnected in lower-dimensional projections [7].

### B. Convex Relaxation of AC-SR

To improve the tractability of AC-SR characterization, many existing studies replace the original non-convex AC-PF equations with convexified formulations, including SOCP-, QCQP-, and SDP-based relaxations [18], [19], [20], [21]. These relaxations introduce auxiliary lifted variables $\boldsymbol{W} \in \mathcal{W}$ through a lifting map $\phi\colon \boldsymbol{Y} \mapsto \boldsymbol{W}$, which abstracts voltage-product nonlinearities into tractable convex form [19], [20]. Accordingly, the original non-convex power flow equations are reformulated in terms of $\boldsymbol{W}$, and the exact lifting consistency $\boldsymbol{W} = \phi(\boldsymbol{Y})$ is relaxed into a convex relaxation set $\mathcal{C}_{\boldsymbol{Y}}$, yielding the following relaxed security-constrained AC-PF feasible set $\mathcal{F}_{\boldsymbol{X},\boldsymbol{Y},\boldsymbol{W}}^{\mathrm{RAC}}$:

$$\mathcal{F}_{\boldsymbol{X},\boldsymbol{Y},\boldsymbol{W}}^{\mathrm{RAC}} \triangleq \left\{ \begin{matrix} (\boldsymbol{X},\boldsymbol{Y},\boldsymbol{W}) \\ \mathcal{X}\times\mathcal{Y}\times\mathcal{C}_{\boldsymbol{Y}} \end{matrix} \middle| \begin{matrix} \boldsymbol{G}_{\mathrm{inj}}(\boldsymbol{X},\boldsymbol{W}) = \boldsymbol{0}, \\ \boldsymbol{H}_{\mathrm{cap}}(\boldsymbol{X},\boldsymbol{W}) \le \boldsymbol{0} \end{matrix} \right\} \tag{10}$$

despite different formulations across various convex relaxation models, these methods share a common lifting-and-relaxation structure that yields tractable outer approximations of the original feasible set $\mathcal{F}_{\boldsymbol{X},\boldsymbol{Y}}^{\mathrm{AC}}$. Consequently, $\mathcal{F}_{\boldsymbol{X},\boldsymbol{Y},\boldsymbol{W}}^{\mathrm{RAC}}$ serves as a computationally tractable convex outer approximation of $\mathcal{F}_{\boldsymbol{X},\boldsymbol{Y}}^{\mathrm{AC}}$, whose projection onto the observed-variable space $\boldsymbol{X}$ defines the corresponding convex-relaxed AC-SR.

**Definition 3** (*Convex-Relaxed AC-SR*): The *convex-relaxed AC-SR* is defined on the observed variables $\boldsymbol{X}$ as the projection of the relaxed AC-PF feasible set, onto $\boldsymbol{X}$:

$$\mathcal{E}_{\mathrm{SR}}^{\mathrm{AC}} \triangleq \left\{\boldsymbol{X} \middle| \exists (\boldsymbol{Y},\boldsymbol{W}) \in \mathcal{Y}\times\mathcal{W}, (\boldsymbol{X},\boldsymbol{Y},\boldsymbol{W}) \in \mathcal{F}_{\boldsymbol{X},\boldsymbol{Y},\boldsymbol{W}}^{\mathrm{RAC}}\right\}, \tag{11}$$

**Proposition 1:** The convex-relaxed AC-SR forms a convex outer approximation of the AC-SR, i.e., $\Omega_{\mathrm{SR}}^{\mathrm{AC}} \subseteq \mathcal{E}_{\mathrm{SR}}^{\mathrm{AC}}$.

*Proof:* The convexity of $\mathcal{E}_{\mathrm{SR}}^{\mathrm{AC}}$ follows directly from the convexity of $\mathcal{F}_{\boldsymbol{X},\boldsymbol{Y},\boldsymbol{W}}^{\mathrm{RAC}}$, as the projection of a convex set is convex. For the inclusion, consider any $\boldsymbol{X} \in \Omega_{\mathrm{SR}}^{\mathrm{AC}}$, By Definition **2**, there exists $\boldsymbol{Y}^{*} \in \mathcal{Y}$ such that $(\boldsymbol{X},\boldsymbol{Y}^{*}) \in \mathcal{F}_{\boldsymbol{X},\boldsymbol{Y}}^{\mathrm{AC}}$. Introducing the convex lifting map $\phi\colon \boldsymbol{Y} \mapsto \boldsymbol{W}$ and noting that the exact lifting consistency $\boldsymbol{W}^{*} = \phi(\boldsymbol{Y}^{*})$ is relaxed into $\boldsymbol{W} \in \mathcal{C}_{\boldsymbol{Y}}$ in $\mathcal{F}_{\boldsymbol{X},\boldsymbol{Y},\boldsymbol{W}}^{\mathrm{RAC}}$, it follows that $(\boldsymbol{X},\boldsymbol{Y},\boldsymbol{W}) \in \mathcal{F}_{\boldsymbol{X},\boldsymbol{Y},\boldsymbol{W}}^{\mathrm{RAC}}$, hence $\boldsymbol{X} \in \mathcal{E}_{\mathrm{SR}}^{\mathrm{AC}}$, thereby establishing $\Omega_{\mathrm{SR}}^{\mathrm{AC}} \subseteq \mathcal{E}_{\mathrm{SR}}^{\mathrm{AC}}$. ■

**Discussion of convex-relaxed AC-SR:** Well-established cut-generation and decomposition approaches [18], [21] can efficiently construct $\mathcal{E}_{\mathrm{SR}}^{\mathrm{AC}}$ as a convex polytope. However, the inclusion $\Omega_{\mathrm{SR}}^{\mathrm{AC}} \subseteq \mathcal{E}_{\mathrm{SR}}^{\mathrm{AC}}$ is generally strict. Consequently, as illustrated in Fig. 1(a), the convex-relaxed AC-SR may admit operating points that are AC-PF insecure under Definition **1**, leading to overestimated security margins that cannot be detected through the convexified model alone.

### C. AC-SR Refinement via Exclusion

As noted above, operating points admitted by $\mathcal{E}_{\mathrm{SR}}^{\mathrm{AC}}$ but violating the exact AC-PF security constraints cannot be distinguished through the convexified model alone. This motivates a *refinement-by-exclusion* strategy: defining the region to be excluded as:

$$\mathcal{I}_{\mathrm{ER}} \triangleq \mathcal{E}_{\mathrm{SR}}^{\mathrm{AC}} \backslash \Omega_{\mathrm{SR}}^{\mathrm{AC}}, \tag{12}$$

and progressively removing it from $\mathcal{E}_{\mathrm{SR}}^{\mathrm{AC}}$ to recover the exact AC-SR $\Omega_{\mathrm{SR}}^{\mathrm{AC}}$, as illustrated in Fig. 1(b). Since $\mathcal{I}_{\mathrm{ER}}$ is implicitly shaped by the nonlinear AC-PF equations and generally non-convex, it cannot be captured by any single simple region. The proposed AC-ITER approach therefore approximates $\mathcal{I}_{\mathrm{ER}}$ through a collection of convex local subregions $\{\mathcal{I}_{\mathrm{ER},k}\}$ whose union progressively covers the AC-PF insecure portion of $\mathcal{E}_{\mathrm{SR}}^{\mathrm{AC}}$. The resulting refinement procedure is developed in Section III.

## III. AC-ITER Approach for AC-SR Refinement

As established in Section II, AC-ITER implements through an iterative *refinement-by-exclusion* strategy operating on a residual region $\mathcal{S}\subseteq\mathcal{E}_{\mathrm{SR}}^{\mathrm{AC}}$, initialized as $\mathcal{S}_0=\mathcal{E}_{\mathrm{SR}}^{\mathrm{AC}}$. As illustrated in

[1] Detailed derivations of the AC-OPF constraints in the compact form $\boldsymbol{AX}+\boldsymbol{BY}-\boldsymbol{D} \le \boldsymbol{0}$ are provided in Appendix A [25].

Fig. 2, AC-ITER refines the convex-relaxed AC-SR $\mathcal{E}^{\mathrm{AC}}_{\mathrm{SR}}$ through a sequence of local exclusion operations. Starting from $\mathcal{S}_0{=}\mathcal{E}^{\mathrm{AC}}_{\mathrm{SR}}$, the algorithm iteratively identifies AC-PF insecure operating points, constructs convex local exclusion subregions around them, and removes the resulting subregions from the current residual region. each iteration consists of three steps:

A) ***Detection Step***: Evaluate the current $\mathcal{S}$ according to the statistical security criterion in Section III-A. If an AC-PF insecure point $\boldsymbol{X}^* \in \mathcal{S}$ is detected, proceed to Step B). otherwise, terminate only when $\mathcal{S}$ satisfies the prescribed statistical security criterion.

B) ***Exclusion Step***: Construct a local convex exclusion subregion around $\boldsymbol{X}^*$ that approximates a portion of $\mathcal{I}_{\mathrm{ER}}$, as detailed in Section III-B. The resulting subregion, denoted as $\mathcal{I}_{\mathrm{ER},k}$ for iteration $k \in \mathfrak{K}$, where $\mathfrak{K}$ denotes the set of all refinement iterations in AC-ITER.

C) ***Refinement Step***: Remove the constructed exclusion subregion $\mathcal{I}_{\mathrm{ER}}$ from $\mathcal{S}_{k-1}$ to obtain a tighter residual region $\mathcal{S}_k$, as detailed in Section III-C.

The procedure repeats until no AC-PF insecure point is detected among a sufficient number of sampled points within $\mathcal{S}$, upon which $\mathcal{S}$ is regarded as satisfying the prescribed security criterion. The resulting refined AC-SR is then given by the final residual region $\mathcal{S}_k$,

$$\mathcal{S}_k = \mathcal{S}_{k-1} \backslash \mathcal{I}_{\mathrm{ER},k} = \mathcal{E}^{\mathrm{AC}}_{\mathrm{SR}} \backslash \textstyle\bigcup_{k\in\mathfrak{K}} \mathcal{I}_{\mathrm{ER},k}. \tag{13}$$

### A. Statistical Security Criterion for AC-SR Refinement

Consider the current residual region $\mathcal{S}{\subseteq}\mathcal{E}^{\mathrm{AC}}_{\mathrm{SR}}$ in the observed-variable space $\mathcal{X}$. To determine whether $\mathcal{S}$ contains AC-PF insecure points, an ideal formulation is to seek the worst-case violation of the AC-PF security constraints over $\mathcal{S}$. Introducing non-negative slack variables $\boldsymbol{S} = \left[\boldsymbol{S}^+_{\mathrm{inj}}, \boldsymbol{S}^-_{\mathrm{inj}}, \boldsymbol{S}^+_{\mathrm{cap}}\right]^{\mathrm{T}} \geq \boldsymbol{0}$ to relax the AC-PF security constraints defining $\mathcal{F}^{\mathrm{AC}}_{\boldsymbol{X},\boldsymbol{Y}}$, this verification can be formulated as the following max-min problem:

$$\Pi^+(\mathcal{S}) = \max_{\boldsymbol{X}\in\mathcal{S}} \ \delta^+(\boldsymbol{X}) \tag{14}$$

$$\begin{aligned} \delta^+(\boldsymbol{X}) = \min_{\boldsymbol{S}\geq\boldsymbol{0},\boldsymbol{Y}} \ & \boldsymbol{1}^{\mathrm{T}}\boldsymbol{S}^+_{\mathrm{inj}} + \boldsymbol{1}^{\mathrm{T}}\boldsymbol{S}^-_{\mathrm{inj}} + \boldsymbol{1}^{\mathrm{T}}\boldsymbol{S}^+_{\mathrm{cap}} \\ \text{s.t.} \ \ & \boldsymbol{G}_{\mathrm{inj}}(\boldsymbol{X},\boldsymbol{Y}) = \boldsymbol{S}^+_{\mathrm{inj}} - \boldsymbol{S}^-_{\mathrm{inj}} : \boldsymbol{\mu}_\alpha \\ & \boldsymbol{H}_{\mathrm{cap}}(\boldsymbol{X},\boldsymbol{Y}) \leq \boldsymbol{S}^+_{\mathrm{cap}} : \boldsymbol{\mu}_\beta \end{aligned} \tag{15}$$

$$\begin{cases} \delta^+(\boldsymbol{X}) = 0 \Leftrightarrow \boldsymbol{X} \in \Omega^{\mathrm{AC}}_{\mathrm{SR}} \\ \delta^+(\boldsymbol{X}) > 0 \Leftrightarrow \boldsymbol{X} \notin \Omega^{\mathrm{AC}}_{\mathrm{SR}} \end{cases}, \begin{cases} \Pi^+(\mathcal{S}) = 0 \Leftrightarrow \mathcal{S} \subseteq \Omega^{\mathrm{AC}}_{\mathrm{SR}} \\ \Pi^+(\mathcal{S}) > 0 \Leftrightarrow \mathcal{S}\backslash\Omega^{\mathrm{AC}}_{\mathrm{SR}} \neq \emptyset \end{cases} \tag{16}$$

where the inner minimization $\delta^+(\boldsymbol{X})$ (15) seeks the minimum relaxation required to satisfy the AC-PF equations and operational security constraints for the given $\boldsymbol{X}$, with dual multipliers $(\boldsymbol{\mu}_\alpha, \boldsymbol{\mu}_\beta)$ retained at optimality[2]. Hence, $\delta^+(\boldsymbol{X}) = 0$ implies that the given $\boldsymbol{X}$ is AC-PF secure, whereas $\delta^+(\boldsymbol{X}) > 0$ certifies that it is AC-PF insecure. The outer maximization $\Pi^+(\mathcal{S})$ (14) extends this pointwise assessment to the entire region $\mathcal{S}$, yielding the equivalences in (16).

**Discussions of** $\Pi^+(\cdot)$: Deterministically certifying $\Pi^+(\mathcal{S}) = 0$ is computationally difficult, as it requires proving the absence of AC-PF-insecure operating points in the residual region $\mathcal{S}$ with respect to the exact nonconvex AC-PF security model.

[2] $\delta^+(\cdot)$ (15) is solved via the interior-point solver IPOPT [23], Due to the nonconvexity of $\mathcal{F}^{\mathrm{AC}}_{\boldsymbol{X},\boldsymbol{Y}}$, only local optimality is guaranteed. Hence, some AC-PF secure points may occasionally be classified as insecure, resulting in a slight under-estimation, which we assume to be acceptable for operators.

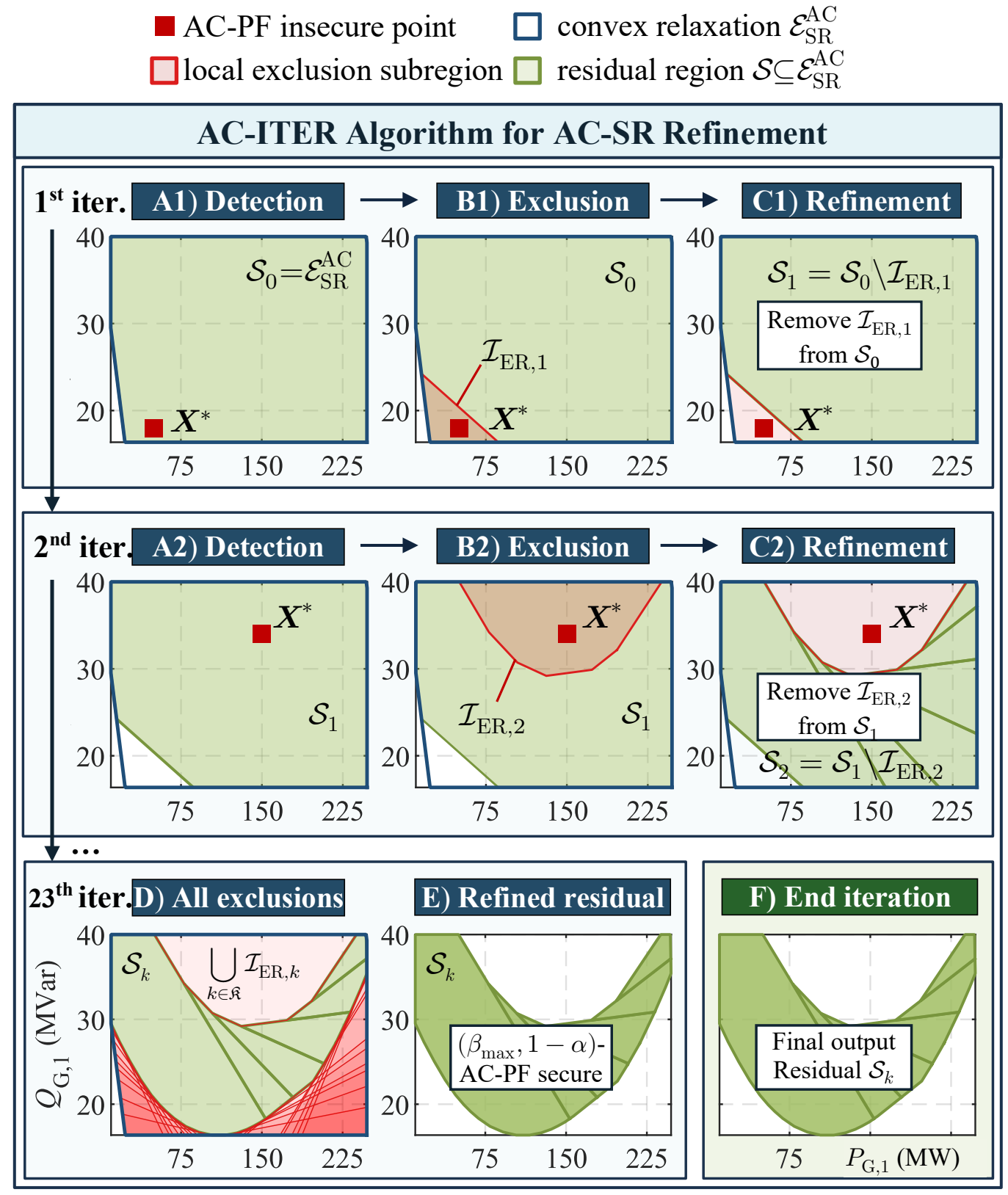

Fig. 2. Illustration of the AC-ITER refinement process for the IEEE 9-bus system projected onto $(P_{\mathrm{G},1}, Q_{\mathrm{G},1})$, with generator-bus voltage limits [1.005,1.015] p.u. and generator reactive power limits [0,40] MVar.

This task is further complicated by the fact that $\mathcal{S}$ may become nonconvex through the ***Exclusion*** and ***Refinement Steps***. Moreover, as noted in Footnote 2, the pointwise evaluation of $\delta^+(\cdot)$ is a nonconvex minimization problem. Embedding such pointwise assessments into a deterministic certificate for $\Pi^+(\mathcal{S}) = 0$ would require resolving $\delta^+(\cdot)$ globally over $\mathcal{S}$, which is beyond what local nonlinear programming solvers can generally guarantee. Conventional bilevel reformulations based on duality, Karush-Kuhn-Tucker conditions, or decomposition do not remove this difficulty, because without convexity or global-optimality guarantees they do not yield a tractable exact certificate for the nonconvex AC feasible set.

Accordingly, this paper turns to the geometric implication of $\Pi^+(\mathcal{S}) = 0$: certifying $\Pi^+(\mathcal{S}) = 0$ is equivalent to verifying $\mathcal{S}{\subseteq}\Omega^{\mathrm{AC}}_{\mathrm{SR}}$. This observation shifts attention from worst-case AC-PF violations to the portion $\mathcal{S}$ lying outside $\Omega^{\mathrm{AC}}_{\mathrm{SR}}$, which is quantified through the following coverage measure.

**Definition 4** (*AC-SR Coverage Deficiency Ratio*): For a candidate region $\mathcal{S}{\subseteq}\mathcal{E}^{\mathrm{AC}}_{\mathrm{SR}}$, the *AC-SR Coverage Deficiency Ratio* (AC-SR CDR) of $\mathcal{S}$ is defined as:

$$\beta(\mathcal{S}) \triangleq \frac{\mu(\mathcal{S}\backslash\Omega^{\mathrm{AC}}_{\mathrm{SR}})}{\mu(\mathcal{S})} \in [0{,}1], \tag{17}$$

where $\mu(\cdot)$ denotes the Lebesgue measure, $\beta(\mathcal{S})$ represents the relative volume of $\mathcal{S}$ lying outside the exact AC-SR. In particular, $\beta(\mathcal{S}) = 0$ is equivalent to $\Pi^+(\mathcal{S}) = 0$, i.e., $\mathcal{S}{\subseteq}\Omega^{\mathrm{AC}}_{\mathrm{SR}}$. Although $\beta(\mathcal{S})$ cannot be computed directly since the geometry of $\Omega^{\mathrm{AC}}_{\mathrm{SR}}$ is unknown, it admits a natural **statistical interpretation**: $\beta(\mathcal{S})$ equals the probability that a point drawn uniformly from $\mathcal{S}$ falls outside $\Omega^{\mathrm{AC}}_{\mathrm{SR}}$. Each uniformly sampled point $\boldsymbol{X}^{(i)} \in \mathcal{S}$, evaluated through $\delta^+(\boldsymbol{X}^{(i)})$, therefore constitutes an independent Bernoulli trial with AC-PF insecure probability $\beta(\mathcal{S})$, reducing intractable worst-case verification (14) to bounding $\beta(\mathcal{S})$ with high confidence through uniform random sampling.

Specifically, when all sampled points within $\mathcal{S}$ are verified AC-PF secure, the accumulated evidence supports the conclusion that $\beta(\mathcal{S})$ is small. Once a sufficient number $N$ of such secure samples has been observed, the region $\mathcal{S}$ can be certified statistically secure with a prescribed confidence level. This observation motivates the following statistical security criterion.

**Definition 5** (*Statistical AC-PF Security Criterion*): Given a tolerated AC-SR CDR $\beta_{\max} \in (0,1)$ and a confidence level $1-\alpha$, the region $\mathcal{S}$ is said to be $(\beta_{\max}, 1-\alpha)$-AC-PF secure if

$$\mathbb{P}[\beta(\mathcal{S}) \leq \beta_{\max}] \geq 1-\alpha, \tag{18}$$

the chance-constrained statistical security criterion relaxes the deterministic requirement $\Pi^+(\mathcal{S})=0$ into a **statistical guarantee**: with confidence $1-\alpha$, no more than a proportion $\beta_{\max}$ of its volume lies outside $\Omega_{\mathrm{SR}}^{\mathrm{AC}}$.

*Role of the Statistical AC-PF Security Criterion in AC-ITER:* The criterion governs the ***Detection Step*** in Fig. 2. The discovery of an AC-PF insecure sample $\boldsymbol{X}^* \in \mathcal{S}$ indicates that the current residual region $\mathcal{S}$ still contains operating points outside $\Omega_{\mathrm{SR}}^{\mathrm{AC}}$ and therefore triggers the ***Exclusion Step***. Otherwise, sampling continues until the prescribed number of AC-PF secure samples required by $(\beta_{\max}, 1-\alpha)$, denoted as $N(\beta_{\max}, 1-\alpha)$, is accumulated within $\mathcal{S}$. Once this threshold is reached, $\mathcal{S}$ is accepted as $(\beta_{\max}, 1-\alpha)$-AC-PF secure and the refinement process in AC-ITER terminates. The required sample size $N(\beta_{\max}, 1-\alpha)$ is derived in Section IV.

### B. Local Exclusion Subregion Construction

When a sampled point $\boldsymbol{X}^* \in \mathcal{S}$ is found AC-PF insecure, the ***Exclusion Step*** seeks to construct a convex exclusion subregion $\mathcal{I}_{\mathrm{ER},k}$ around $\boldsymbol{X}^*$ approximating a portion of $\mathcal{I}_{\mathrm{ER}}$. To this end, this subsection develops a *detect-and-refine* strategy that iteratively detects SR boundary points and generates the corresponding local exclusion cuts (LECs), where each LEC is a supporting half-space of $\Omega_{\mathrm{SR}}^{\mathrm{AC}}$ whose outer side contains AC-PF insecure operating points. The intersection of the accumulated LECs forms a convex polytope enclosing $\boldsymbol{X}^*$, yielding the exclusion subregion $\mathcal{I}_{\mathrm{ER},k}$ to be removed from $\mathcal{S}$.

B1) **AC-SR boundary point detection**. The search starts from $\mathcal{D}^{(0)} = \mathcal{E}_{\mathrm{SR}}^{\mathrm{AC}}$. At each iteration $i$, the nearest AC-PF secure point to $\boldsymbol{X}^*$ within the current search region $\mathcal{D}^{(i)}$ is obtained through the feasibility-distance projection:

$$\Delta\boldsymbol{X}^{\dagger} = \underset{\Delta\boldsymbol{X},\boldsymbol{Y}}{\operatorname{argmin}} \left\{ \|\Delta\boldsymbol{X}\|_2^2 \colon \mathrm{s.t.} \begin{array}{c} (\boldsymbol{X}^* + \Delta\boldsymbol{X}, \boldsymbol{Y}) \in \mathcal{F}_{\boldsymbol{X},\boldsymbol{Y}}^{\mathrm{AC}} \\ \boldsymbol{X}^* + \Delta\boldsymbol{X} \in \mathcal{D}^{(i)} \end{array} \right\},$$
$$\boldsymbol{X}_{(i)}^{\dagger} = \boldsymbol{X}^* + \Delta\boldsymbol{X}^{\dagger}, \tag{19}$$

if (19) is infeasible, no AC-PF secure point remains in $\mathcal{D}^{(i)}$ and the *detect-and-refine* strategy terminates. Otherwise, $\boldsymbol{X}_{(i)}^{\dagger}$ satisfies the exact AC-PF security constraints and is therefore AC-PF secure, i.e., $\delta^+\left(\boldsymbol{X}_{(i)}^{\dagger}\right) = 0$. Since $\boldsymbol{X}^* \notin \Omega_{\mathrm{SR}}^{\mathrm{AC}}$, the minimum-distance point $\boldsymbol{X}_{(i)}^{\dagger}$ must lie on the boundary of $\Omega_{\mathrm{SR}}^{\mathrm{AC}}$, and any interior point would admit a strictly closer AC-PF secure point along the segment toward $\boldsymbol{X}^*$, contradicting optimality. Thus $\boldsymbol{X}_{(i)}^{\dagger}$ is a certified boundary point of $\Omega_{\mathrm{SR}}^{\mathrm{AC}}$.

B2) **LEC construction at the boundary point**. Having identified the certified AC-SR boundary point $\boldsymbol{X}_{(i)}^{\dagger}$, the next step is to extract local boundary information of $\Omega_{\mathrm{SR}}^{\mathrm{AC}}$ around it. Since $\delta^+\left(\boldsymbol{X}_{(i)}^{\dagger}\right) = 0$, the local variation of $\delta^+(\cdot)$ provides a first-order description of the boundary geometry. The gradient of $\delta^+(\cdot)$ with respect to $\boldsymbol{X}$ at $\boldsymbol{X}_{(i)}^{\dagger}$ can be obtained from the dual multipliers $(\boldsymbol{\mu}_{\alpha}, \boldsymbol{\mu}_{\beta})$ associated with (15):

$$\nabla_{\boldsymbol{X}}\delta^+\left(\boldsymbol{X}_{(i)}^{\dagger}\right) = \left.\frac{\partial\delta^+(\boldsymbol{X})}{\partial\boldsymbol{X}}\right|_{\boldsymbol{X}=\boldsymbol{X}_{(i)}^{\dagger}} = \boldsymbol{\mu}_{\alpha}^{\mathrm{T}}\frac{\partial\boldsymbol{G}_{\mathrm{inj}}}{\partial\boldsymbol{X}} + \boldsymbol{\mu}_{\beta}^{\mathrm{T}}\frac{\partial\boldsymbol{H}_{\mathrm{cap}}}{\partial\boldsymbol{X}}, \tag{20}$$

The gradient points toward directions along which $\delta^+(\cdot)$ increases, and therefore locally indicates the side of the boundary corresponding to AC-PF insecure operating points. The corresponding LEC retains this side[3]:

$$\mathrm{LEC}\left(\boldsymbol{X}_{(i)}^{\dagger}\right) = \left\{\boldsymbol{X}\colon \left(\frac{\nabla_{\boldsymbol{X}}\delta^+(\boldsymbol{X}_{(i)}^{\dagger})}{\|\nabla_{\boldsymbol{X}}\delta^+(\boldsymbol{X}_{(i)}^{\dagger})\|_2}\right)^{\mathrm{T}} \left(\boldsymbol{X} - \boldsymbol{X}_{(i)}^{\dagger}\right) \geq 0\right\}, \tag{21}$$

the validity of the constructed LEC is established in Appendix B [25]. In particular, the generated cut always encloses the triggering AC-PF insecure point $\boldsymbol{X}^*$, while locally separating the AC-SR boundary at the detected boundary point $\boldsymbol{X}_{(i)}^{\dagger}$.

B3) **Iterative refinement of the exclusion subregion**. A single LEC provides only a local linear approximation of the boundary of $\mathcal{I}_{\mathrm{ER}}$ and generally cannot isolate the entire $\mathcal{I}_{\mathrm{ER}}$ surrounding $\boldsymbol{X}^*$. Therefore, after constructing $\mathrm{LEC}\left(\boldsymbol{X}_{(i)}^{\dagger}\right)$, the search region $\mathcal{D}^{(i)}$ is updated and Step a) is repeated to identify an additional boundary point $\boldsymbol{X}_{(i+1)}^{\dagger}$ and generate a new LEC.

To avoid repeatedly detecting the same boundary point, the search region $\mathcal{D}^{(i)}$ is contracted by a resolution margin $\epsilon > 0$:

$$\mathcal{D}^{(i+1)} \leftarrow \mathcal{D}^{(i)} \bigcap \left(\left(\frac{\nabla_{\boldsymbol{X}}\delta^+(\boldsymbol{X}_{(i)}^{\dagger})}{\|\nabla_{\boldsymbol{X}}\delta^+(\boldsymbol{X}_{(i)}^{\dagger})\|_2}\right)^{\mathrm{T}} \left(\boldsymbol{X} - \boldsymbol{X}_{(i)}^{\dagger}\right) \geq \epsilon\right), \tag{22}$$

The *detect-and-refine* strategy alternates between boundary-point detection and LEC generation until (19) becomes infeasible. Let $\mathfrak{J}_k$ denote the index set of all LECs generated during this process. At termination, the accumulated LECs define the local exclusion subregion

$$\mathcal{I}_{\mathrm{ER},k}(\boldsymbol{X}^*) = \mathcal{E}_{\mathrm{SR}}^{\mathrm{AC}} \bigcap \left(\bigcap\nolimits_{i\in\mathfrak{J}_k} \mathrm{LEC}\left(\boldsymbol{X}_{(i)}^{\dagger}\right)\right), \tag{23}$$

which encloses $\boldsymbol{X}^*$ and is the object removed from $\mathcal{S}_k$.

**Discussions of** $\epsilon > 0$: A strictly positive $\epsilon > 0$ is introduced to prevent repeated detection of previously identified boundary points and ensure finite termination. Since $\epsilon$ only affects the contraction of the search region $\mathcal{D}^{(i)}$ and does not modify the LECs, the resulting exclusion subregion $\mathcal{I}_{\mathrm{ER},k}(\boldsymbol{X}^*)$ may conservatively absorb a small portion of AC-SR, particularly near locally concave boundary regions. This approximation increases with $\epsilon$. As $\epsilon$ decreases, the approximation error diminishes, and $\mathcal{I}_{\mathrm{ER},k}(\boldsymbol{X}^*)$ more closely approaches the local portion of $\mathcal{I}_{\mathrm{ER}}$. However, smaller values of $\epsilon$ require more refinement iterations, while the limiting case $\epsilon = 0$ would generally lead to an infinite search process and is therefore not used in practice.

### C. Iterative Exclusion Refinement

Section III-B constructs an exclusion subregion $\mathcal{I}_{\mathrm{ER},k}$ around detected AC-PF insecure point $\boldsymbol{X}^*$. The ***Refinement Step*** then removes this $\mathcal{I}_{\mathrm{ER},k}$ from the current residual region $\mathcal{S}_k$, yielding

[3] Appendix B [25] establishes the boundary and local approximation properties and gives a multiplier reselection rule ensuring containment of $\boldsymbol{X}^*$ under the stated conditions.

$$\mathcal{S}_k = \mathcal{S}_{k-1} \backslash \mathcal{I}_{\mathrm{ER},k}(\boldsymbol{X}^*), \beta(\mathcal{S}_k) \triangleq \frac{\mu(\mathcal{S}_k \backslash \Omega_{\mathrm{SR}}^{\mathrm{AC}})}{\mu(\mathcal{S}_k)} \leq \beta(\mathcal{S}_{k-1}), \quad (24)$$

thereby reducing the AC-SR CDR of $\mathcal{S}_k$ does not exceed that of $\mathcal{S}_{k-1}$. After multiple exclusions, $\mathcal{S}_k$ may become nonconvex or disconnected, making direct uniform sampling from $\mathcal{S}_k$ difficult. To generate candidate points for the next ***Detection Step*** without explicitly constructing $\mathcal{S}_k$, AC-ITER represents the accumulated exclusion subregions $\{\mathcal{I}_{\mathrm{ER},k}\}_{k\in\mathfrak{K}}$ through the following logical constraints[4]:

$$\underbrace{\bigwedge_{k\in\mathfrak{K}} \left( \bigvee_{j\in\mathfrak{J}_k} \left( \left( \frac{\nabla_{\boldsymbol{X}^\dagger}\delta^+}{\|\nabla_{\boldsymbol{X}^\dagger}\delta^+\|_2} \right)^{\mathrm{T}} \left( \boldsymbol{X} - \boldsymbol{X}^\dagger_{(i)} \right) < 0 \right) \right)}_{\text{equivalent to } \boldsymbol{X} \notin \mathcal{I}_{\mathrm{ER},k}(\boldsymbol{X}^*)}, \quad (25)$$

which compactly enforce $\boldsymbol{X} \notin \bigcup_{k\in\mathfrak{K}} \mathcal{I}_{\mathrm{ER},k}$, with $\mathfrak{K}$ the index set of excluded subregions and $\mathfrak{J}_k$ the LECs of the $k$-th subregion. Candidate points are sampled independently and uniformly from $\mathcal{E}_{\mathrm{SR}}^{\mathrm{AC}}$, and the logical constraints in (25) are used to reject those lying in any previously excluded subregion. Consequently, the retained candidates are independently and uniformly distributed over $\mathcal{S}_k$ and are passed to the next ***Detection Step*** for pointwise AC-PF security assessment.

## IV. Statistical Termination and Computational Acceleration for AC-ITER

Section III defines the ***Detection Step*** of AC-ITER as uniform sampling from the current $\mathcal{S}_k$ followed by pointwise AC-PF security assessment $\delta^+(\cdot)$. Section IV-A derives the minimum sample size required for termination when every sampled point is verified as AC-PF secure, Section IV-B develops the AC-SSA approach to accelerate repeated pointwise assessments without changing the security criterion, and Section IV-C integrates both components into the practical ***Detection Step***.

### *A. Finite-Sample Guarantees for A Fixed Residual Region*

As established in Section III.A, AC-ITER terminates when the residual region $\mathcal{S}$ is certified as $(\beta_{\max}, 1-\alpha)$-AC-PF secure, which requires accumulating a sufficient number of AC-PF secure samples within $\mathcal{S}$. Since each pointwise security outcome $\{\boldsymbol{X}^{(1)}, \boldsymbol{X}^{(2)}, \ldots, \boldsymbol{X}^{(N)}\}$ constitutes an independent Bernoulli trial with AC-PF insecure probability $\beta(\mathcal{S})$, the empirical frequency of insecure points converges to $\beta(\mathcal{S})$ as $N \to \infty$. Under finite sampling, however, the goal shifts to certifying $\beta(\mathcal{S}) \leq \beta_{\max}$ with confidence $1-\alpha$. This subsection derives the required sample size $N(\beta_{\max}, 1-\alpha)$ explicitly.

**Proposition 2** (*Finite-Sample Certification of $(\beta_{\max}, 1-\alpha)$-AC-PF security*): If $N$ points are independently and uniformly sampled from a fixed region $\mathcal{S}$ and every point is verified as AC-PF secure, i.e., $\delta^+(\{\boldsymbol{X}^{(1)}, \boldsymbol{X}^{(2)}, \ldots, \boldsymbol{X}^{(N)}\}) = 0$, then $\mathcal{S}$ is $(\beta_{\max}, 1-\alpha)$-AC-PF secure in the sense of Definition **5**, provided that:

$$N \geq \left\lceil \frac{\ln \alpha}{\ln(1-\beta_{\max})} \right\rceil \quad (26)$$

*Proof:* Under the null hypothesis $\beta(\mathcal{S}) > \beta_{\max}$, the probability that all Bernoulli trials $\delta^+(\{\boldsymbol{X}^{(1)}, \boldsymbol{X}^{(2)}, \ldots, \boldsymbol{X}^{(N)}\}) = 0$ yield AC-PF secure outcomes is at most $(1-\beta_{\max})^N$. To guarantee that this probability does not exceed the significance level $\alpha$, it suffices to require $(1-\beta_{\max})^N \leq \alpha$, and solving gives $N \geq \frac{\ln \alpha}{\ln(1-\beta_{\max})}$. Therefore, observing all $N$ points AC-PF secure, the hypothesis $\beta(\mathcal{S}) > \beta_{\max}$ can be rejected with confidence level $1-\alpha$, the false-acceptance probability is at most $\alpha$. ■

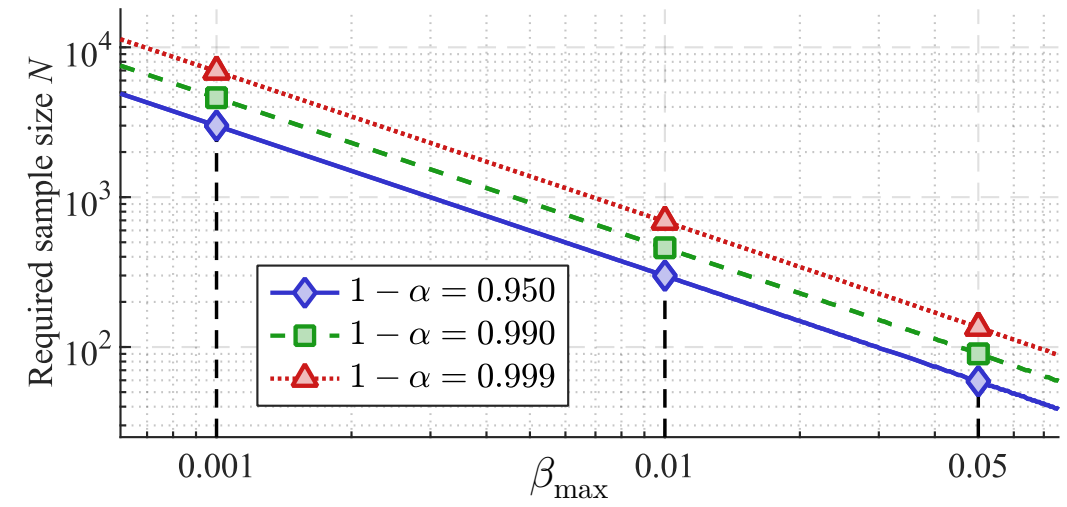

Fig. 3. Minimum sample size $N(\beta_{\max}, 1-\alpha)$ required for statistical certification when all sampled points are verified as AC-PF secure, under different tolerated $\beta_{\max}$ and confidence levels $1-\alpha$, as given by Proposition **2**.

TABLE I
Minimum Sample Size Required When All Points Are AC-PF Secure

| $N(\beta_{\max}, 1-\alpha)$ | $\beta_{\max}=0.050$ | $\beta_{\max}=0.010$ | $\beta_{\max}=0.001$ |
|---|---|---|---|
| $1-\alpha=0.950$ | 59 | 299 | 2995 |
| $1-\alpha=0.990$ | 90 | 459 | 4603 |
| $1-\alpha=0.999$ | 135 | 688 | 6905 |

Proposition **2** converts the termination condition of the ***Detection Step*** into an all-secure sample threshold. (26) shows that this sample size increases as $\beta_{\max}$ decreases or confidence level $1-\alpha$ increases, as illustrated in Fig. 3 and Table I. For example, under the setting $\beta_{\max}=0.001$ and $1-\alpha=0.999$, the minimum all-secure sample size is 6905.

In AC-ITER, however, each detected AC-PF insecure sample triggers an exclusion that updates $\mathcal{S}_{k-1}$ to $\mathcal{S}_k$. For the practical implementation, we adopt a retained-sample reuse strategy. During each ***Detection Step***, previously verified AC-PF secure samples that satisfy the updated logical constraints in (25) are first retained. Additional samples are then independently and uniformly drawn from the current residual region and evaluated pointwise for AC-PF security. This strategy avoids restarting the sample count after every refinement.

**Discussion of retained-sample reuse strategy**: By allowing the same sampled security outcomes to support both residual $\mathcal{S}_k$ refinement and termination, this strategy reduces repeated sampling and AC-PF security assessments but comes at the cost of independent validation within AC-ITER. Consequently, the fixed-region confidence statement in Proposition **2** does not carry over exactly to AC-ITER as a whole, although the spatial uniformity of the retained samples is preserved. Despite this limitation, the case study in Section V show that the resulting residual regions achieve high empirical accuracy, supporting the use of (26) as a practical termination threshold.

### *B. AC-SSA Acceleration*

Proposition **2** specifies the all-secure sample threshold $N$ for terminating the ***Detection Step***. A direct implementation evaluates $\delta^+(\{\boldsymbol{X}^{(1)}, \boldsymbol{X}^{(2)}, \ldots, \boldsymbol{X}^{(N)}\})$ for every newly sampled point by solving the nonconvex program with IPOPT. When the current residual region $\mathcal{S}_k$ has a small AC-SR CDR, most sampled points are secure. Although retained-sample reuse avoids restarting the sample count after refinement, newly sampled points still require repeated IPOPT evaluations.

As reported in [22], neighboring AC-OPF solutions often retain the same *active-variable* set (variables at their upper or lower bounds) and *active-constraint* sets (binding inequality constraints) under moderate operating perturbations, suggesting that solution information obtained from $\delta^+(\cdot)$ can be reused to assess nearby points. Building on this observation and the parallel *hypothesis-then-verify* strategy proposed in [2], we develop the AC-PF security similarity-based acceleration (AC-SSA) approach for the ***Detection Step***.

[4] The implicit logical representation is used throughout the refinement. For visualization, the final residual region $\mathcal{S}_k$ can be reconstructed as an explicit union of polytopes via the MPT3 toolbox, as shown in the case studies figures.

By reusing the *active-variable* and *active-constraint* sets associated with previously verified AC-PF-secure samples, AC-SSA attempts to verify new candidates as AC-PF secure before invoking IPOPT, thereby reducing repeated full IPOPT solves. The AC-SSA approach consists of the following steps.

a0) **Initialization:** To initialize AC-SSA, $N_{\text{init}}$ points are first independently and uniformly sampled from $\mathcal{S}_0$ and evaluated pointwise $\delta^+(\cdot)$ in sampling order using IPOPT. Each detected AC-PF-insecure point triggers the corresponding ***Exclusion*** and ***Refinement Steps*** in Section III, the remaining unevaluated points and verified AC-PF-secure samples are then re-filtered using (25), and this process continues until every point in the initial batch has either been evaluated or removed. The retained AC-PF-secure samples form the initial reference set for AC-SSA, whose size is denoted by $N_0$. For each reference $\boldsymbol{X}^{(i)}$, $\delta^+(\boldsymbol{X}^{(i)}) = 0$ implies $\boldsymbol{S}^+_{\text{inj}}, \boldsymbol{S}^-_{\text{inj}}, \boldsymbol{S}^+_{\text{cap}} = \boldsymbol{0}$ at optimality. From the corresponding IPOPT solution, AC-SSA records the *active-variable* set $\boldsymbol{Y}^{(i)}_{\text{act}}$, *inactive-variables* $\boldsymbol{Y}^{(i)}_{\text{inact}}$ (variables not at their bounds), and the associated *active constraint* set.

a1) **AC-PF Security Similarity Assumption:** For each newly sampled point $\boldsymbol{X}$ in ***Detection Step***, AC-SSA selects the reference $\boldsymbol{X}^{(i)}$ that minimizes $\|\boldsymbol{X} - \boldsymbol{X}^{(i)}\|_2$ and defines $\Delta\boldsymbol{X} = \boldsymbol{X} - \boldsymbol{X}^{(i)}$. It assumes that the *active-variables* and *constraint*s identified from $\delta^+(\boldsymbol{X}^{(i)})$ remain active for $\boldsymbol{X}$. AC-SSA then fixes $\boldsymbol{Y}^{(i)}_{\text{act}}$ at its reference value and seeks the unknown $\hat{\boldsymbol{Y}}_{\text{inact}}$ from the reduced nonlinear system $\boldsymbol{K}^{(i)}(\cdot,\cdot)$:

$$\hat{\boldsymbol{Y}}_{\text{inact}} \text{ is sol. of } \underbrace{\boldsymbol{K}^{(i)}\left(\boldsymbol{X}^{(i)} + \Delta\boldsymbol{X}, \boldsymbol{Y}^{(i)}_{\text{act}}\right)}_{2n+c \text{ constraints, } 4n+2g-a \text{ variables}} = \boldsymbol{0} \tag{27}$$

$$\boldsymbol{K}^{(i)}\left(\boldsymbol{X}, \left[\boldsymbol{Y}^{(i)}_{\text{act}}, \hat{\boldsymbol{Y}}_{\text{inact}}\right]\right) = \begin{bmatrix} \boldsymbol{G}_{\text{inj}}\left(\boldsymbol{X}, \left[\boldsymbol{Y}^{(i)}_{\text{act}}, \hat{\boldsymbol{Y}}_{\text{inact}}\right]\right) \\ \underbrace{\boldsymbol{H}^{(i)}_{\text{act}}\left(\boldsymbol{X}, \left[\boldsymbol{Y}^{(i)}_{\text{act}}, \hat{\boldsymbol{Y}}_{\text{inact}}\right]\right)}_{\text{active constraints}} \end{bmatrix} \tag{28}$$

where $c$ is the number of inherited active inequality constraints included in $\boldsymbol{H}^{(i)}_{\text{act}}$, and $a$ denotes the number of variables fixed by the identified *active-variable* set. In (27)-(28), $\boldsymbol{X} = \boldsymbol{X}^{(i)} + \Delta\boldsymbol{X}$ and $\boldsymbol{Y}^{(i)}_{\text{act}}$ are treated as fixed inputs of $\boldsymbol{K}^{(i)}(\cdot)$. Specifically, the blue-marked quantities ($\boldsymbol{Y}^{(i)}_{\text{act}}$, $\boldsymbol{H}^{(i)}_{\text{act}}(\cdot,\cdot)$, and $\boldsymbol{K}^{(i)}(\cdot,\cdot)$) are inherited from the reference solution of $\delta^+(\boldsymbol{X}^{(i)})$. $\hat{\boldsymbol{Y}}_{\text{inact}}$ is the only unknown vector. Their combination $\left[\boldsymbol{Y}^{(i)}_{\text{act}}, \hat{\boldsymbol{Y}}_{\text{inact}}\right]$ forms a candidate solution of $\delta^+\left(\boldsymbol{X}^{(i)} + \Delta\boldsymbol{X}\right)$.

a2) **Candidate Solution Recovery and Verification:** Because (27) may be over- or underdetermined, AC-SSA updates $\hat{\boldsymbol{Y}}_{\text{inact}}$ using the pseudoinverse iteration in (29). At each iteration $j$, the current $\hat{\boldsymbol{Y}}_{\text{inact},j}$ is placed at its original indices $\mathcal{N}_j$ and combined with the variables fixed in $\mathcal{A}_j$ to reconstruct the full state $\hat{\boldsymbol{Y}}_j$. Since the pseudoinverse update does not enforce variable bounds, some *inactive-variables* $Y_{j,m}$ may move outside their feasible ranges ($\mathcal{B}^+_j$ or $\mathcal{B}^-_j$), the corresponding variables are fixed at their bounds and transferred to $\mathcal{A}_{j+1}$, where they remain fixed thereafter. The next update is applied only to the remaining *inactive variables* $\mathcal{N}_{j+1}$.

$$\begin{aligned}
&\Delta\hat{\boldsymbol{Y}}_{\mathcal{N}_j,j} \leftarrow -\boldsymbol{J}^\dagger_j \boldsymbol{K}^{(i)}(\boldsymbol{X}^{(i)} + \Delta\boldsymbol{X}, \boldsymbol{Y}_j),\\
&\boldsymbol{J}^\dagger_j = \left(\partial \boldsymbol{K}^{(i)}(\boldsymbol{X}^{(i)} + \Delta\boldsymbol{X}, \boldsymbol{Y}_j)/\partial\hat{\boldsymbol{Y}}_{\mathcal{N}_j}\right)^\dagger,\\
&\hat{\boldsymbol{Y}}_{\mathcal{N}_j,j+1} \leftarrow \hat{\boldsymbol{Y}}_{\mathcal{N}_j,j} + \Delta\hat{\boldsymbol{Y}}_{\mathcal{N}_j,j}, \hat{\boldsymbol{Y}}_{\mathcal{A}_j,j+1} \leftarrow \hat{\boldsymbol{Y}}_{\mathcal{A}_j,j},\\
&\mathcal{B}^+_j = \{m \in \mathcal{N}_j\colon Y_{j+1,m} \geq \overline{Y}_m\}, \mathcal{B}^-_j = \{m \in \mathcal{N}_j\colon Y_{j+1,m} < \underline{Y}_m\},\\
&\begin{cases} \mathcal{A}_{j+1} \leftarrow \mathcal{A}_j \bigcup \mathcal{B}^+_j \bigcup \mathcal{B}^-_j\,, \\ \mathcal{N}_{j+1} \leftarrow \mathcal{U}_Y \backslash \mathcal{A}_{j+1}, \end{cases}, \hat{Y}_{j+1,m} \leftarrow \begin{cases} \overline{Y}_m, m \in \mathcal{B}^+_j, \\ \underline{Y}_m, m \in \mathcal{B}^-_j, \\ \hat{Y}_{j+1,m}, \text{ otherwise}, \end{cases}
\end{aligned} \tag{29}$$

where $(\cdot)^\dagger$ denotes the Moore-Penrose pseudoinverse, $\mathcal{U}_Y$ denotes the index set of all variables in $\hat{\boldsymbol{Y}}$, $\mathcal{N}_j = \mathcal{U}_Y \backslash \mathcal{A}_j$ is the current *inactive-variable* index set. If the iteration fails to converge, the candidate remains unresolved. Otherwise, the converged state $\hat{\boldsymbol{Y}}_j$ is substituted into the AC-PF constraints in (30).

$$\begin{cases} \boldsymbol{G}_{\text{inj}}(\boldsymbol{X}^{(i)} + \Delta\boldsymbol{X}, \hat{\boldsymbol{Y}}_j) = \boldsymbol{0} \\ \boldsymbol{H}_{\text{cap}}(\boldsymbol{X}^{(i)} + \Delta\boldsymbol{X}, \hat{\boldsymbol{Y}}_j) \leq \boldsymbol{0} \end{cases}, \tag{30}$$

a candidate $\boldsymbol{X}$ is verified as AC-PF secure in the sense of Definition **1** without invoking IPOPT only if all original constraints in (30) are satisfied; otherwise, it remains unresolved.

a3) **Parallel Evaluation:** The *hypothesis-then-verify* strategy in A1)-A2) can be performed independently and distributed across available CPU cores using parallel constructs such as MATLAB `parfor` construct. This parallel structure allows AC-SSA to attempt preliminary verification of new samples without invoking IPOPT, thereby improving the computational efficiency of the ***Detection Step***.

### C. AC-SSA-Assisted Detection Step

Let $N = N(\beta_{\max}, 1-\alpha)$. With the retained-sample reuse strategy in Section IV-A and the AC-SSA a0) initialization step in Section IV-B, the ***Detection Step*** is implemented as follows.

A1) **Sample Update:** Previously verified AC-PF-secure samples and the remaining unevaluated points are re-filtered using (25), while their relative sampling order is preserved. Let $n_k$ denote the number of remaining points. If $n_k < N$, $2N - n_k$ additional points are independently and uniformly sampled from $\mathcal{S}_k$ and appended in sampling order. The additional points serve as a conservative reserve for subsequent refinements.

A2) **Parallel Assessment:** AC-SSA evaluates all newly sampled points in parallel according to steps a1)-a3) in Section IV-B. Points satisfying (30) are verified as AC-PF secure, while the AC-PF security of the remaining points remains unknown.

A3) **Detection or Termination:** Following the preserved sampling order, only points whose AC-PF security remains unknown are evaluated pointwise $\delta^+(\cdot)$ using IPOPT. If an AC-PF-insecure point is identified, it is passed to the ***Exclusion Step***. If $N$ points are verified as AC-PF secure first, the *Detection Step* terminates. If the current point set is exhausted before either condition is met, the procedure returns to A1).

## V. Case Study

The proposed AC-ITER and AC-SSA approaches are evaluated on the IEEE 9-, 118-bus systems and a regional transmission benchmark derived from a practical network in Shandong Province, China, hereafter referred to as SD-507. To protect confidentiality, its topology, equipment identifiers, and operating data have been anonymized and modified. SD-507 consists of 507 buses, 739 transmission branches, and 134 generators. All test systems use the most stringent setting in Table I, with $\beta_{\max} = 0.001$ and $1 - \alpha = 0.999$. Equation (26) gives a minimum all-secure sample size of $N = 6905$ as the AC-ITER termination threshold.

To quantify the AC-SR characterization accuracy of different methods, 100,000 new points are independently and uniformly sampled from the convex-relaxed region $\mathcal{E}_{\text{SR}}^{\text{AC}}$ and assessed pointwise by solving $\delta^{+}(\cdot)$. Based on these assessments, the following sample-based estimates are computed:

- *Estimated AC-SR CDR*: For each candidate region $\mathcal{S}$, its estimated AC-SR CDR, denoted by $\hat{\beta}(\mathcal{S})$, is the fraction of AC-PF-insecure samples among those lying in $\mathcal{S}$.
- *Estimated Secure-Point Removal Ratio*: For each candidate region $\mathcal{S}$, its estimated secure-point removal ratio, denoted by $\hat{\gamma}(\mathcal{S})$, is the fraction of AC-PF-secure samples lying outside $\mathcal{S}$ among all AC-PF-secure samples.

All nonlinear optimization problems associated with the AC-PF model are solved in IPOPT 3.11.7 [23] with the MA57 linear solver. All experiments are conducted on a PC equipped with an Intel Core Ultra 7 265K CPU at 3.90 GHz and 48 GB of RAM, using MATLAB R2024b with 20 parallel workers. MPT3 is used only to generate the final $\mathcal{S}_k$ visualizations.

### A. AC-SR Geometry and Method Comparison

As discussed in Section I, simplified geometric representations may not fully capture AC-SRs with nonconvex boundaries and internal holes. This subsection first compares three methods on the IEEE 9-bus system. **M1** is the QCQP-based convex-relaxed SR with McCormick envelopes [20], constructed using the method in [18], and corresponds to the initial convex-relaxed region $\mathcal{E}_{\text{SR}}^{\text{AC}}$. **M2** is the inscribed-hypercube approximation [24], whereas **M3** is the final residual region $\mathcal{S}_k$ obtained by AC-ITER. The AC-SR geometries of the IEEE 118-bus and SD-507 systems are then examined under different loading conditions and voltage-limit settings.

For the IEEE 9-bus system, two choices of the observed-variable vector $\boldsymbol{X}$ and two generator-voltage settings are considered. Cases A and B use the active and reactive power injections of the generator connected to bus 1, $(P_{\text{G},1}, P_{\text{G},2})$, and the active power injections of the generators connected to buses 1 and 2, $(P_{\text{G},1}, Q_{\text{G},1})$, respectively. Setting I constrains generator-bus voltage limits of [1.005,1.015] p.u., whereas Setting II fixes them at 1.01 p.u. Both settings use the same generator reactive-power limits of [0,40] MVar. Unless otherwise stated, $\epsilon = 0.005$ is used for AC-ITER in all four cases, Case-9I-A, 9II-A, 9I-B, and 9II-B and summarized in Table II.

Fig. 4(a) compares the geometries obtained by the three methods and shows that the $\Omega_{\text{SR}}^{\text{AC}}$ geometry changes with both the observed-variable selection and generator-voltage setting. Cases A and B exhibit different projected shapes, while Settings I and II further alter the feasible range and boundary shape under the same observed variables. Table II reports the corresponding estimates $\hat{\beta}(\cdot)$ and $\hat{\gamma}(\cdot)$. As an outer approximation, **M1** cannot capture the internal nonconvex features of the AC-SR and therefore includes many AC-PF-insecure points, with $\hat{\beta}_{\mathbf{M1}}$ ranging from 11.922% to 79.689%. **M2** provides a secure inner approximation but removes a large fraction of secure points, with $\hat{\gamma}_{\mathbf{M2}}$ ranging from 69.654% to 89.976%. In comparison, **M3** removes local exclusion subregions from the initial $\mathcal{E}_{\text{SR}}^{\text{AC}}$, reducing $\hat{\beta}_{\mathbf{M3}}$ to 0.001%-0.290% and $\hat{\gamma}_{\mathbf{M3}}$ to 0.414%-1.861%. Overall, **M3** provides a closer geometric characterization of $\Omega_{\text{SR}}^{\text{AC}}$ by reducing the insecure portions retained by **M1** while avoiding the excessive removal of secure points by **M2**.

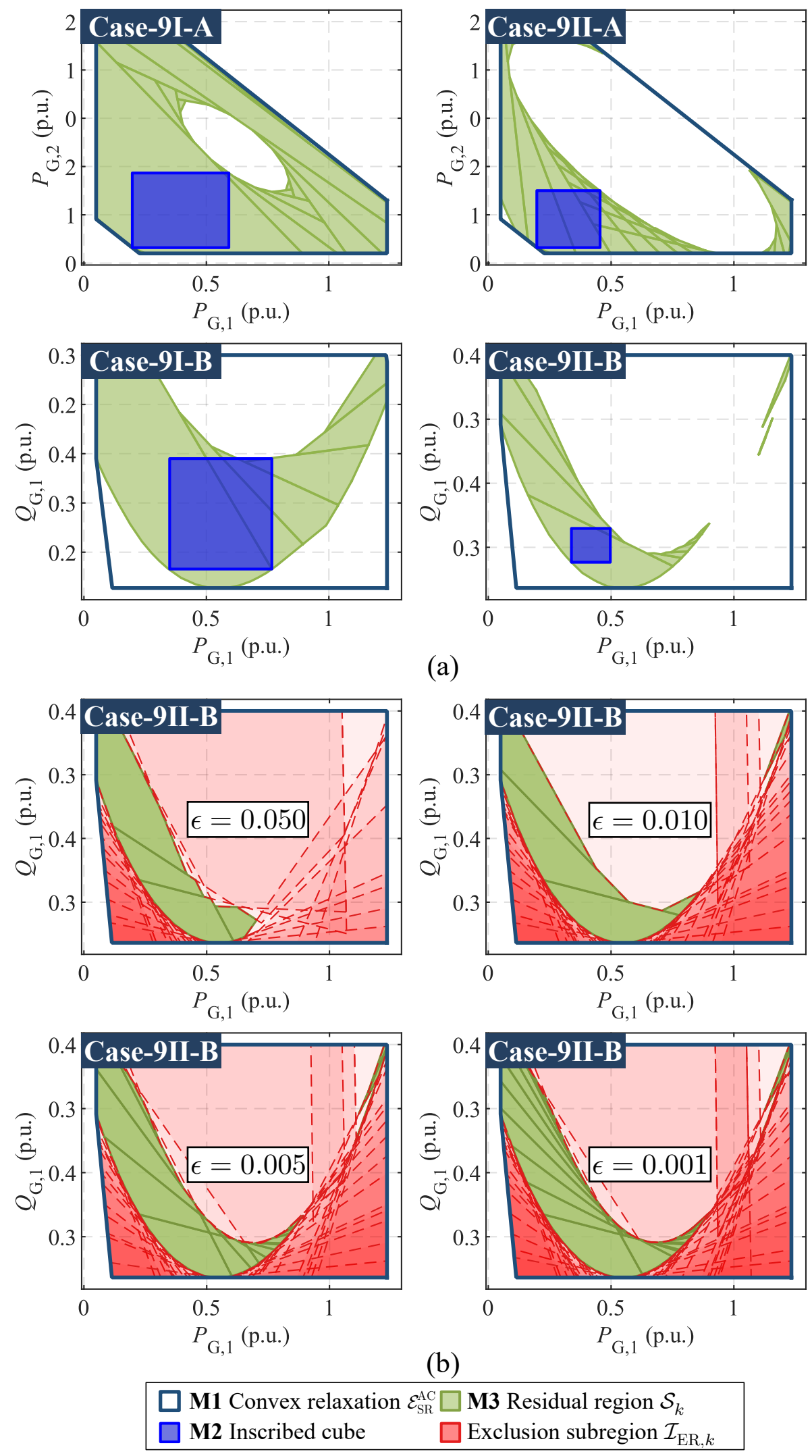

Fig. 4. Geometric refinement of the AC-SR on the IEEE 9-bus system. (a) Comparison of **M1**, **M2**, and **M3** across four configurations. (b) Sensitivity of the AC-ITER to the contraction margin parameter $\epsilon$.

TABLE II
CONFIGURATIONS AND GEOMETRICS RESULTS OF THE IEEE 9-BUS SYSTEM

| Result | Case-9I-A | Case-9II-A | Case-9I-B | Case-9II-B |
|---|---|---|---|---|
| Number of LECs | 20 | 38 | 28 | 52 |
| Exclusion subregions | 5 | 10 | 23 | 35 |
| $\hat{\beta}$ of **M1** (%) | 11.922 | 63.936 | 44.579 | 79.689 |
| $\hat{\beta}$ of **M3** (%) | **0.001** | **0.020** | **0.120** | **0.290** |
| $\hat{\gamma}$ of **M2** (%) | 82.181 | 78.047 | 69.654 | 89.976 |
| $\hat{\gamma}$ of **M3** (%) | **0.414** | **0.757** | **1.777** | **1.861** |

Three-dimensional AC-SRs are further examined on the IEEE 118-bus and SD-507 systems to assess the effects of loading levels and voltage-security limits on AC-SR geometry, respectively. For the IEEE 118-bus system, the observed vector comprises the reactive power injections of the generators at buses 92, 99, and 103, i.e., $\boldsymbol{X} = (Q_{\text{G},92}, Q_{\text{G},99}, Q_{\text{G},103})$. Case-118I and Case-118II use the standard and doubled active and reactive loads, respectively, with the PV-bus voltage magnitudes fixed at the corresponding AC-OPF values. For SD-507, the observed vector comprises the active and reactive power injections of the generator at bus 5 and the reactive power injection of the generator at bus 4, i.e., $\boldsymbol{X} = (P_{\mathrm{G},4}, P_{\mathrm{G},5}, Q_{\mathrm{G},5})$. Case-SDI and Case-SDII impose generator-bus voltage-magnitude limits of [0.850,1.150] and [0.875,1.125] p.u., respectively.

TABLE III
SUMMARY OF AC-ITER REFINEMENT RESULTS

| Result | Case-118I | Case-118II | Case-SDI | Case-SDII |
|---|---|---|---|---|
| Exclusion time (s) | 9.42 | 3.21 | 58.50 | 14.29 |
| AC-ITER time (s) | 21.04 | 22.65 | 175.32 | 185.53 |
| $\beta(\mathcal{S}_k)$ (%) | 0.18 | 0.24 | 0.22 | 0.31 |
| $\hat{\gamma}(\mathcal{S}_k)$ (%) | 0.12 | 0.18 | 3.45 | 4.22 |
| Exclusion subregions | 46 | 23 | 29 | 28 |

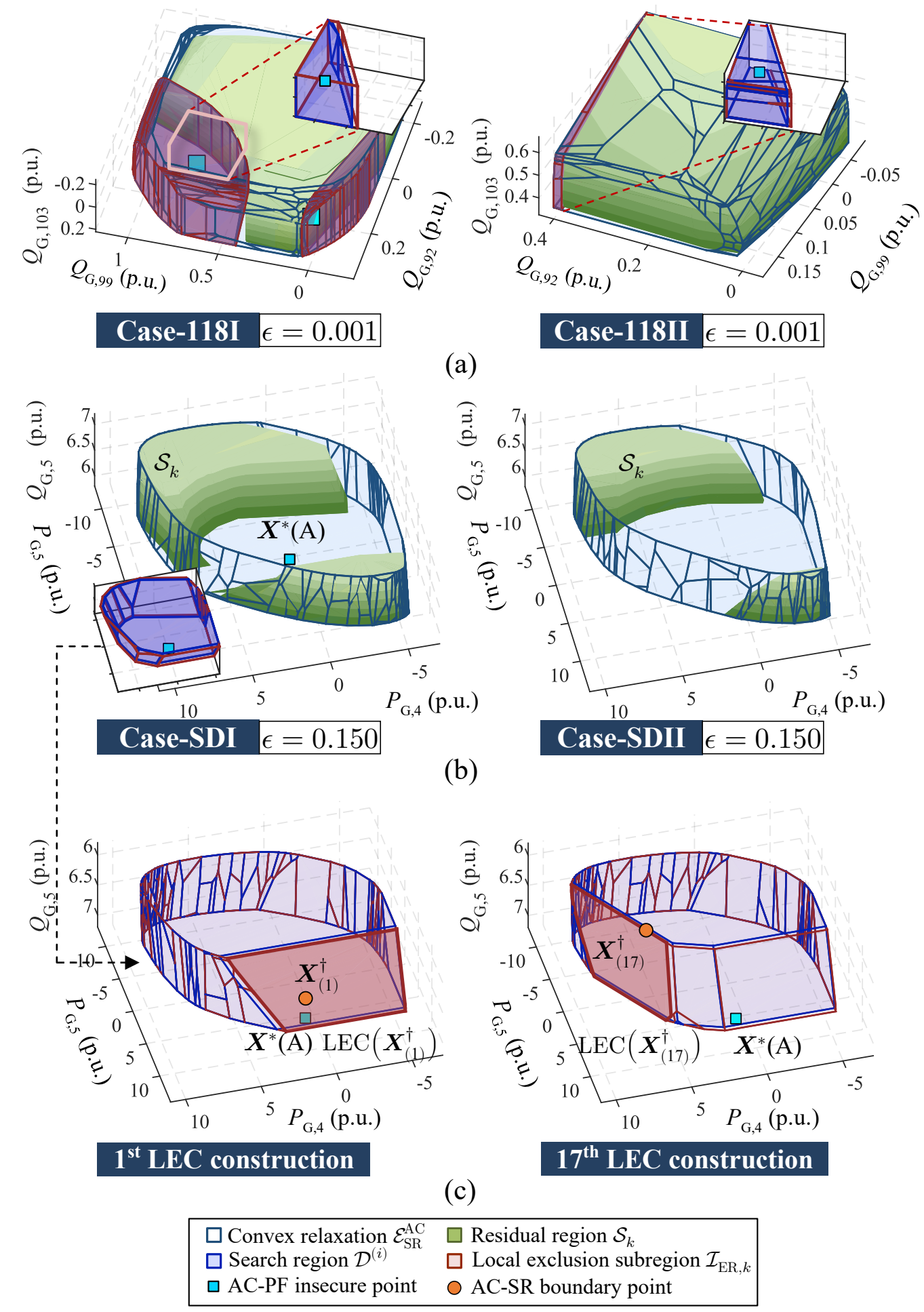

Fig. 5. Comparison of M1 and M3 on the IEEE 118-bus and SD-507 systems. (a) Case-118I and Case-118II. (b) Case-SDI and Case-SDII. (c) LEC construction and search-region contraction at $\boldsymbol{X}^*(\mathrm{A})$ in Case-SDI.

Fig. 5 (a) and (b) compare **M1** and **M3** in these four cases. For the IEEE 118-bus system, doubling the loads changes both the range and boundary shape of the AC-SR. For SD-507, tightening the generator-bus voltage limits causes a less pronounced change, while the overall nonconvex geometry remains similar. Together, these results show that nonconvex boundaries and internal holes can arise across the tested systems and operating conditions. Overall, **M3** reduces the insecure portion of **M1** and the excessive conservativeness of **M2**, providing a more accurate characterization of the AC-SR.

### B. AC-ITER Refinement Performance

*1) Results Across Test Cases:* Table II and Table III quantify

TABLE IV
EFFECT OF THE CONTRACTION MARGIN $\epsilon$ ON AC-ITER IN CASE-9II-B

| Result | $\epsilon = 0.050$ | $\epsilon = 0.010$ | $\epsilon = 0.005$ | $\epsilon = 0.001$ |
|---|---|---|---|---|
| Number of LECs | 33 | 44 | 52 | 92 |
| Exclusion subregions | 26 | 35 | 35 | 41 |
| Residual polytopes | 4 | 6 | 23 | 53 |
| $\beta$ of **M3** (%) | 0.14 | 0.26 | 0.29 | 0.23 |
| $\hat{\gamma}$ of **M3** (%) | 13.74 | 5.41 | 1.86 | 0.71 |
| AC-ITER Time (s) | 5.10 | 5.68 | 6.21 | 7.95 |

the refinement accuracy of AC-ITER. For the IEEE 9-bus cases, **M3** reduces the insecure portions retained by **M1** while preserving substantially more secure points than **M2**. For the IEEE 118-bus and SD-507 cases, the estimated $\hat{\beta}(\mathcal{S})$ ranges from 0.18% to 0.31%, with secure-point removal ratios of 0.12%-4.22%. These results show that AC-ITER removes insecure portions while retaining most secure operating points among compared methods.

*2) Iterative Refinement Process:* Fig. 2 illustrates successive exclusions from the residual region, whereas Fig. 5(c) shows the first and seventeenth LEC constructions around the detected AC-PF-insecure point A in Case-SDI. At each illustrated iteration, an AC-SR boundary point is obtained by solving (19), and the corresponding LEC is constructed using (21). The accumulated LECs progressively refine the local exclusion subregion containing A, while the inward contraction in (22) updates the search region for subsequent boundary-point detection.

*3) Effect of the Contraction Margin*: Fig. 4(b) and Table IV examine the sensitivity of AC-ITER to $\epsilon$ in Case-9II-B. As $\epsilon$ decreases from 0.050 to 0.001, additional LECs provide a finer representation of the curved AC-PF boundary, reducing the secure-point exclusion rate from 13.74% to 0.71%. Meanwhile, the numbers of LECs, exclusion subregions, and residual polytopes increase from 33, 26, and 4 to 92, 41, and 53, respectively, and the computation time rises from 5.10 to 7.95. The residual insecure rate remains below 0.3% for all settings. Overall, $\epsilon = 0.005$ offers a favorable accuracy-efficiency tradeoff for AC-ITER in this case.

### C. AC-SSA Acceleration Performance

This subsection evaluates the acceleration provided by AC-SSA for the pointwise AC-PF evaluations in the ***Detection Step*** of AC-ITER. Since this benefit is more relevant when each evaluation of $\delta^+(\cdot)$ requires greater computational effort, the IEEE 118-bus and SD-507 systems are selected. Accordingly, **DS1** and **DS2** are compared using identical ordered sample sequences, termination settings, and solver settings.

- **DS1** ($N_{\mathrm{init}} = 0$): AC-SSA is disabled, and uniformly sampled points are evaluated $\delta^+(\cdot)$ in sampling order.
- **DS2** ($N_{\mathrm{init}} > 0$): The proposed AC-SSA-assisted ***Detection Step*** is used, with $N_{\mathrm{init}}$ set to 100, 250, 500, and 1000.

Fig. 6(a) compares the computation time required by the ***Detection Step*** under **DS1** and **DS2** with different $N_{\mathrm{init}}$ settings. For **DS2**, the computation time is divided into AC-SSA initialization, AC-SSA verification, and pointwise IPOPT evaluation of samples whose AC-PF security remains unknown. Fig. 6(b) shows the resulting initial reference-set size $N_0$ and the reduction in ***Detection-Step*** IPOPT calls relative to **DS1**. Increasing $N_{\mathrm{init}}$ enlarges $N_0$ may provide more reference solutions for subsequent AC-SSA verification but also increases the initialization effort; consequently, the most efficient setting depends

TABLE V
COMPUTATION PERFORMANCE OF **DS1** AND **DS2** IN THE DETECTION STEP

| Case | Case-118I | Case-118II | Case-SDI | Case-SDII |
|---|---|---|---|---|
| $N_{\text{init}}^*$ | 100 | 100 | 1000 | 1000 |
| $N_0^*$ | 76 | 94 | 339 | 165 |
| **DS1** Time (s) | 112.57 | 127.60 | 364.17 | 354.79 |
| **DS2**$^*$ Time (s) | 10.71 | 11.67 | 74.83 | 86.16 |
| **DS1** IPOPT Calls | 6952 | 6928 | 6934 | 6942 |
| **DS2**$^*$ IPOPT Calls | 359 | 346 | 1255 | 1307 |

Notation: $N_{\text{init}}^*$ denotes the tested initialization size yielding the shortest **DS2** evaluation time among $N_{\text{init}} \in \{100,250,500,1000\}$.

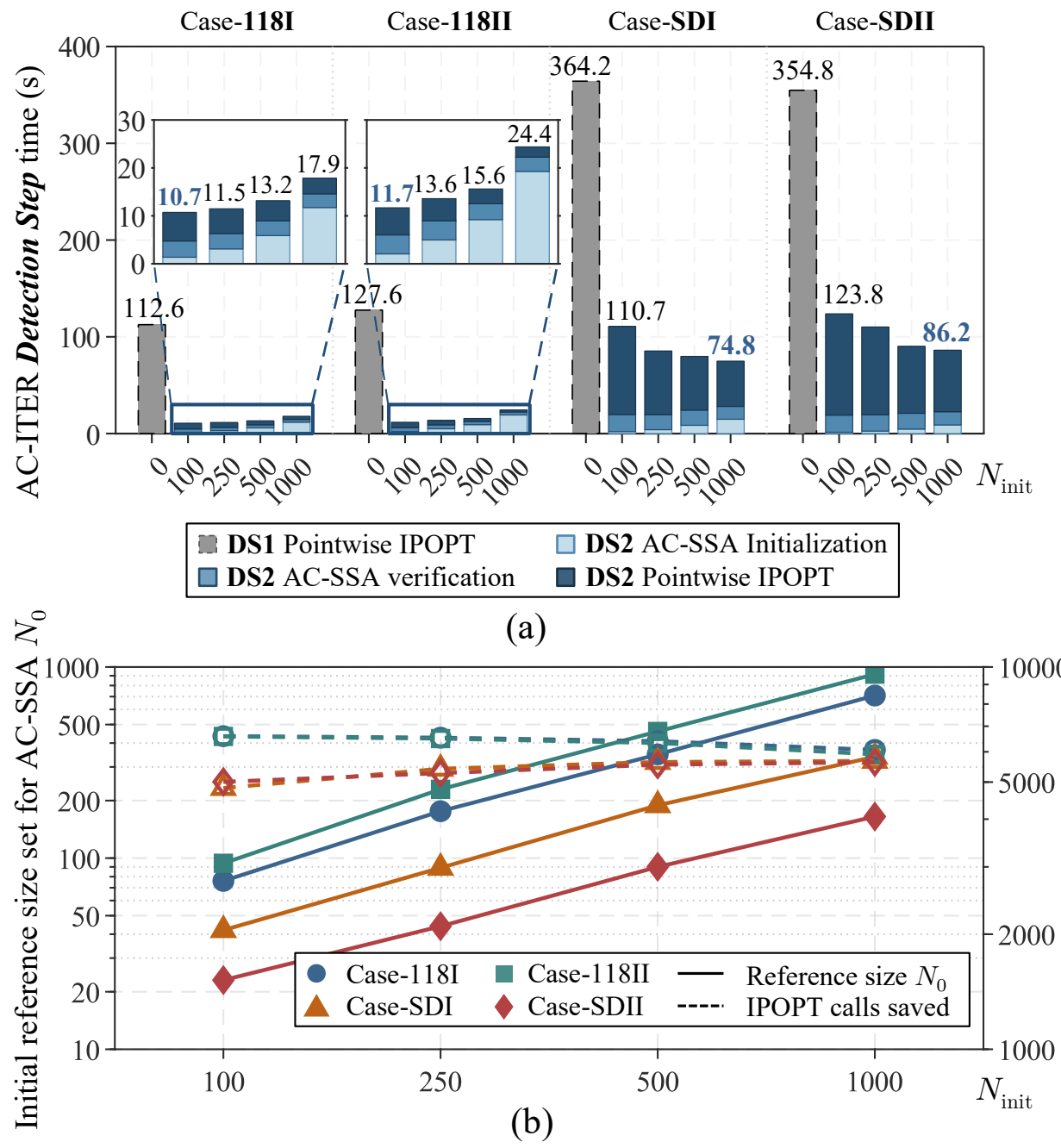

Fig. 6. Effect of AC-SSA Initialization Size on Detection-Step Performance

on the relative computational cost of AC-SSA verification and IPOPT evaluation. Table V summarizes the best observed **DS2** results among the tested $N_{\text{init}}$ settings. The shortest **DS2** evaluation time is obtained at $N_{\text{init}} = 100$ for both IEEE 118-bus cases and at $N_{\text{init}} = 1000$ for Case-SDI and Case-SDII, producing $N_0 = 76$, 94, 339, and 165, respectively. These evaluation times represent 9.15%-24.28% of the **DS1** times, while the IPOPT call reductions account for 81.17%-95.01% of the **DS1** totals. These results demonstrate that AC-SSA effectively improves the computational efficiency of the ***Detection Step*** in AC-ITER by reusing nearby verified secure solutions to replace most repeated pointwise IPOPT evaluations with lower-cost verification.

## VI. CONCLUSION

This paper proposes AC-ITER to refine convex-relaxed AC security regions through successive local exclusions. Boundary-point searches and dual-based LECs construct exclusion subregions, while AC-SSA reduces repeated nonlinear optimization calls by reusing verified secure solutions. Tests on the IEEE 9- and 118-bus systems and SD-507 demonstrate reduced retention of insecure points, limited removal of secure points, and faster security assessment. The resulting regions constitute numerical refinements rather than exact global characterizations. Future work will strengthen exclusion validity and statistical guarantees under adaptive sampling.